\documentclass[aps,pra,showpacs,twocolumn,groupedaddress]{revtex4}

\usepackage{hyperref}
\usepackage{graphicx}
\usepackage{amsmath}
\usepackage{amsfonts}
\usepackage{amssymb}
\usepackage{epsfig}
\usepackage{bm}

\begin{document}

\title{Rayleigh-Taylor instability in binary condensates}

\author{S. Gautam and D. Angom}
\affiliation{Physical Research Laboratory,
         Navarangpura, Ahmedabad - 380 009\\}

\date{\today}


\begin{abstract}

   We propose a scheme to initiate and examine Rayleigh-Taylor instability in
the two species Bose-Einstein condensates. We identify $^{85}$Rb-$^{87}$Rb 
mixture as an excellent candidate to observe it experimentally. The 
instability is initiated by tuning the $^{85}$Rb-$^{85}$Rb interaction 
through magnetic Feshbach resonance. We show that the observable signature of 
the instability is the damping of the radial oscillation. This would perhaps 
be one of the best controlled experiments on Rayleigh-Taylor instability. 
We also propose a semi analytic scheme to determinate stationary state
of binary condensates with the Thomas-Fermi approximation for the  axis 
symmetric traps. 

\end{abstract}

\pacs{03.75.Mn, 03.75.Kk, }

\maketitle


{\em Introduction}.---Rayleigh-Taylor instability (RTI) sets in when lighter 
fluid supports a heavier one. It is present across a wide spectrum of phenomena 
related to interface of two fluids. The turbulent mixing in 
astrophysics, inertial confinement fusion and geophysics originate from
RTI. In superfluids, RTI sets up crystallization waves at the superfluid-solid 
$^4$He interface \cite{Burmistrov}. Despite the ubiquitous nature and 
importance, controlled experiments with RTI are difficult and rare. 
However, we show that the two species Bose-Einstein condensates (TBECs) or 
binary condensates are ideal systems for a controlled study of RTI in 
superfluids. The remarkable feature of TBECs, absent in single component BECs, 
is the phenomenon of phase separation. The TBECs, first realized in a mixture 
of two hyperfine states of $^{87}$Rb \cite{Myatt}, are rich systems to explore 
nonlinear phenomena. Several theoretical works have examined various aspects 
of TBECs. These include stationary states \cite{Ho,Pu1,Trippenbach,Ao},
modulational instability \cite{Kasamatsu1,Raju,Ronen}, collective excitations 
\cite{Graham,Pu2,Gordon,Svidzinsky} and domain walls solitons\cite{Coen}.  
Another instability related to RTI, which has attracted growing interest, 
is the Kelvin Helmholtz instability (KHI). The prerequisites of KHI are, phase 
separation and relative tangential velocities at the interface. Quantum KHI 
have been observed in experiments with $^3$He \cite{Blaau} and recently 
studied theoretically for TBEC \cite{Takeuchi}.

 To initiate RTI we start with the phase separated state. Then, increase the 
scattering length of the species at the core.  At a certain value it creates 
a quantum analogue of RTI in fluid dynamics. As a case study we choose the 
TBEC of $^{85}$Rb-$^{87}$Rb mixture. In this system, the $^{85}$Rb intra 
species interaction is tunable through a Feshbach resonance \cite{Roberts} and
was recently used to study the miscibility \cite{Papp}.  More recently, the 
dynamical pattern formation during the growth of this  TBEC was theoretically 
investigated \cite{Ronen}. The other feature is, the inter species 
$^{85}$Rb-$^{87}$Rb interaction is also tunable and well studied \cite{Papp2}. 
Considering the parameters of the experimental realization, we choose the 
axis symmetric ( cigar shaped) trap geometry.


{\em Phase separated cigar shaped TBECs}.---In the mean field approximation, 
the TBEC is described by a set of coupled Gross-Pitaevskii equations
\begin{equation}
  \left[ \frac{-\hbar^2}{2m_i}\nabla^2 + V_i(\rho,z) + 
         \sum_{j=1}^2U_{ij}|\psi_j|^2 \right]\psi_{i}(\rho,z) =
         \mu_i\psi_{i}(\rho,z), 
  \label{eq.gp}
\end{equation}
where $i = 1, 2$ is the species index, $U_{ii} = 4\pi\hbar^2a_i/m_i$ 
with $m_i$ as mass and $a_i$ as s-wave scattering length, is the intra-species 
interaction; $U_{ij}=2\pi\hbar^2a_{ij}/m_{ij}$ with $m_{ij}=m_i m_j/(m_i+m_j)$ 
as reduced mass and $a_{ij}$ as inter-species scattering length, is 
inter-species interaction and $\mu_i$ is the chemical potential of the 
$i^{\rm th}$ species. To study the RTI we consider the phase separated state 
($ U_{12}>\sqrt{U_{11}U_{22}}$) in axis symmetric trapping potentials 
$ V_i(\rho,z) = m_i\omega^2(\alpha_i^2\rho^2 + \lambda_i^2 z^2)/2$.
In the present work we consider cigar shaped potentials, that is the 
anisotropy parameters $\alpha_i > \lambda_i$ and $U_{ij}$ are all positive. 
Neglecting the inter species overlap, the Thomas-Fermi (TF) solutions are 
$ |\psi_i(\rho,z)|^2 = [\mu_i-V_i(\rho,z)]/U_{ii}$. The chemical 
potentials $\mu_i$ are fixed through the normalization conditions. When  
$\alpha_i \gg \lambda_i$, the interface of the phase separated state
is planar and species having larger scattering length sandwiches the other 
one \cite{asymm}. 

  For simplicity of analysis consider trapping potentials with coincident 
centers. Then, let $z=\pm L_1$ be the planes separating the two components 
and $\pm L_2$, the spatial extent of the outer species along
$z$-axis. The density distributions $n_1$ and $n_2$ of the TBEC are 
\begin{eqnarray} 
  n_1(\rho, z) &= &\frac{\mu_1 - V_1(\rho, z)}{U_{11}},\,\,\,-L_1<z<L_1, \\
  n_2(\rho, z) &= &\frac{\mu_2 - V_2(\rho, z)}{U_{22}},\,\,\,
                   L_1< |z| <L_2.  
\end{eqnarray} 
This assumes no overlap between the two species.  Then the problem to determine
the stationary state is equivalent to calculating $L_1$. Theoretically, $L_1$ 
can be determined by minimising the total energy of the TBEC with fixed number 
of particles of each species. If $N_i$ and $\rho_i$ are the number of atoms 
and radial size of $i^{\rm th}$ species respectively, then 
\begin{eqnarray}
  N_i = 2\pi\int_0^{\rho_i}\rho d\rho \int_{-L_i}^{L_i}dz |\psi_i(\rho,z)|^2.
  \label{eq.ni}
\end{eqnarray}
From the TF approximation
\begin{eqnarray}
 N_1 &=& \frac{\pi L_1(3\omega^2L_1^4m_1\lambda_1^4-20L_1^2\lambda_1^2\mu_1-
         60(\omega^2m_1-2)\mu_1^2))}{30U_{11}\alpha_1^2},
        \label{eq.mu1} \\
 N_2 &=& \frac{2\pi}{3\lambda_2U_{22}}\left[\frac{L_1^2\lambda_2^2}{
         20\alpha_2^2}(5\omega^2\lambda_2^3L_1^3m_2 -
         8\omega^2m_2(L_1^2\lambda_2^2)^{3/2}\right.  \nonumber\\
     & &-60\lambda_2L_1(\omega^2m_2-1)\mu_2+40(\omega^2m_2-1)L_1
         \lambda_2\mu_2)\nonumber\\
     & &-\frac{\mu_2}{5\alpha_2^2}(-5\omega^2\lambda_2^3L_1^3m_2-15
            \lambda_2L_1(\omega^2m_2-2)\mu_2\nonumber\\
     & &+\left.4\sqrt{2}(3\omega^2m_2-5)\mu_2^{3/2}\right]  .
        \label{eq.mu2} 
\end{eqnarray}
The total energy of the binary condensate is
\begin{eqnarray}
   E  & = & \int dV\left[ V_1(\rho,z)| \psi_1(\rho,z)|^2 +
            V_2(\rho,z)|\psi_2(\rho,z)|^2 + \right . \nonumber  \\
      &   & \left.\frac{1}{2}U_{11}|\psi_1(\rho,z)|^4 + \frac{1}{2}U_{22}|
            \psi_2(\rho,z)|^4 \right] .
   \label{eq.en}
\end{eqnarray}
 We minimize $E$ numerically, with Eq.(\ref{eq.mu1}) and (\ref{eq.mu2}) as 
constraints, to obtain the required value of $L_1$.
Substituting the value of $L_1$ back into Eqs.(\ref{eq.mu1}) and (\ref{eq.mu2}),
one can determine $\mu_1$ and $\mu_2$. Thus Eqs.(\ref{eq.mu1}-\ref{eq.en}) 
uniquely define the stationary state of the TBEC. 

\begin{figure}[h]
   \includegraphics[height=8.5cm,angle=-90]{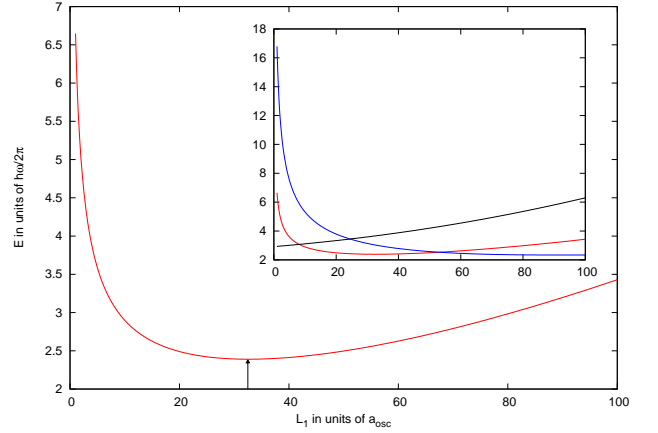}
   \caption{The variation in energy $E$ with $L_1$ in phase separated regime. 
            The upward arrow indicates the position of minimum $E$, which is 
            at $L_1=32.5a_{\rm osc}$. Inset shows the same plot along with 
            the variation of $\mu_1$ and $\mu_2$ with respect to $L_1$,
            the blue and black curves correspond to $\mu_1$ and $\mu_2$ 
            respectively.}
   \label{fig.st_tbec}  
\end{figure}

  As mentioned earlier, we consider the parameters of the recent experiment 
\cite{Papp} with $^{85}$Rb and $^{87}$Rb as the first and second atomic 
species. The radial trapping frequencies are identical ($\alpha_i=1$) and 
for the axial trapping frequencies $\lambda_1=0.022$ and $\lambda_2=0.020$. 
The scattering lengths are $a_{11}=51a_0$, $a_{22}=99a_0$ and 
$a_{12}=a_{21}=214a_0$, and we take $N_i=50,000$. Then, Fig.\ref{fig.st_tbec} 
shows the variation in $E$ as a function of $L_1$. The value of $L_1$
where minimum of $E$ occurs is $32.5a_{\rm osc}$. Here  the unit of length
$a_{\rm osc}=\sqrt{\hbar/m_1\omega}$ with $\omega=130$Hz, is the radial 
trapping frequency. This is in agreement with the numerical result
$33.8a_{\rm osc}$ calculated using split-step Crank-Nicholson 
method (imaginary time propagation) \cite{Muruganandam}. 
We refer to this state as phase I, where $^{85}$Rb and $^{87}$Rb are at the 
center and flanks respectively.
We have also calculated the equations of interface planes for 
trapping potentials whose minima do not coincide. The expressions are much 
more complicated, however the numerical and semi-analytic results are in 
agreement.


{\em Binary condensate evolution}.---In the fluid dynamics parlance, the 
gradient of the trapping potentials are the equivalent of gravity.  If $s$ is 
the oscillation frequency of the interface between the two condensates, one 
placed over the other. Then from Bernoulli's principle along with proper 
boundary conditions \cite{Drazin,Chandrasekhar}, we find from linear stability 
analysis
\begin{equation}
 s = \pm\left[ \frac{\sqrt{k_x^2+k_y^2} m\omega^2\lambda^2L (
     n_1-n_2)}{n_1 + n_2}\right] ^{1/2}
 \label{eq.rt_inst}
\end{equation}
Here $k_x$ and $k_y$ are wave numbers along $x$ and $y$ coordinates. The 
densities $n_1$ and $n_2$ are at a point $(\rho, L)$ on the interface.
For the sake of simplicity, we consider $m_1=m_2=m$ and 
$\lambda_1=\lambda_2=\lambda$ while deriving the above relation. 
There is an instability, referred to as Rayleigh-Taylor instability,
at the interface when $n_1 < n_2$. From the TF approximation this condition
is equivalent to $a_{11}>a_{22}(\mu_1-V)/(\mu_2-V)$. Here $V$ is the trapping 
potential of the two species at the interface. Normal fluids with RTI, any 
perturbation at the interface however small grows exponentially. Then the 
lighter fluid rises to the top as bubbles and heavier fluid sinks as finger 
like extensions till the entire bulk of the lighter fluid is on top of the 
denser one. On the other hand, binary condensates in a similar situation 
evolve in a very different way. 

  To examine the dynamical evolution of the binary condensate with RTI, we 
take phase I ( $a_{11} < a_{22}$) as the initial state. In this phase,
the $^{87}$Rb BEC at the flanks is considered as resting over the 
$^{85}$Rb BEC at the core. Then through the $^{85}$Rb--$^{85}$Rb magnetic 
Feshbach resonance \cite{Roberts} increase $a_{11}$ till 
$a_{11} > a_{22}(\mu_1 - V)/(\mu_2-V)$ to set up RTI. However, maintain 
$ U_{12}>\sqrt{U_{11}U_{22}}$ so that the TBEC is still immiscible. Let us 
call this as the phase Ia and it is an unstable state. The stationary state 
of the new parameters is phase separated and similar in structure to the 
initial state. But with the species interchanged. Let us call the stationary 
state of the new parameters as phase II. The binary condensate should 
dynamically evolve from phase Ia to II. However, unlike in normal fluids 
with RTI,  there are no bulk flows of either $^{85}$Rb or $^{87}$Rb atoms, 
to the periphery of the trap. Instead the condensates tunnel with modulations. 
This occurs due to the coherence in the quantum liquids. To examine the 
evolution, we solve the pair of time-dependent GP equations 
\begin{equation}
  i\hbar\frac{\partial \psi_{i}(\rho,z)}{\partial t}= \left[ 
    \frac{-\hbar^2}{2m_i}\nabla^2 + V_i(\rho,z) + \sum_{j=1}^2U_{ij}|\psi_j|^2 
    \right]\psi_{i}(\rho,z) , 
  \label{eq.tdgp}
\end{equation}
which describe the TBEC. During the evolution, the density profiles is 
approximated as 
$n_i(\rho, z) = n_i^{\rm eq}(\rho, z) + \delta n_i(\rho, z)$.  Here 
$n_i^{\rm eq}(\rho, z)$  and $ \delta n_i(\rho, z)$ are the equilibrium 
density and fluctuation arising from the increase in $a_{11} $. Following 
the hydrodynamic approximations, the $ \delta n_i(\rho, z)$ or collective 
modes follow the equations
\begin{equation}
   m_i\frac{\partial ^2}{\partial t^2}\delta n_i  =  \bm{\nabla} n_i \cdot 
      \bm{\nabla} \sum_{j=1}^2U_{ij}\delta n_j + 
      n_i \nabla ^2\sum_{j=1}^2 U_{ij}\delta n_i .
\end{equation}
Consider $\delta n_i (\rho, z, t) = a_i(t)\rho ^l \exp(\pm il\phi)$ as the 
form of the solution, where $a_i(t)$ subsumes the time dependent part of the 
solution including temporal variation of the amplitude and $l$ is an integer. 
Then as $\nabla^2\delta n_i = 0$ and for the miscible phase, considered for
simplicity of the boundary conditions, we get 
\begin{equation}
  \ddot{a}_i = -\frac{l\omega ^2}{U_{ii}} 
                \left ( U_{ii}a_{11} + U_{ij}a_{22} \right ). 
\end{equation}
We can also get a similar set of coupled equations for the other form of the 
collective modes 
$\delta n_i (\rho, z, t) = a_i(t)z\rho ^{l-1} \exp(\pm i(l-1)\phi) $. In this
case the prefactor is $(l-1+\lambda_i^2) $ instead of $l$. 
In either of the cases, the equations are similar to two coupled  oscillators. 
For the phase separated state, the form of the TF solutions are significantly 
different from the miscible one. However, when RTI sets in, the collective 
modes like in miscible case, are damped and coupled as the condensates 
interpenetrate each other.


{\em TBEC evolution with RTI}.---To examine the evolution of TBEC with RTI, 
as mentioned earlier, we choose the phase I as the initial state. Then change 
$a_{11}$ to $80a_0$, $102a_0,200a_0, 306a_0, 408a_0$ and $780 a_0$, the last 
value is in the miscible parameter region. The dynamical variables which are 
coarse grained representative of the dynamical evolution are 
$\rho_{\rm rms}$ and $z_{\rm rms}$, the {\em rms} radial and axial sizes. 
\begin{figure}[h]
   \includegraphics[height=8cm,angle=-90]{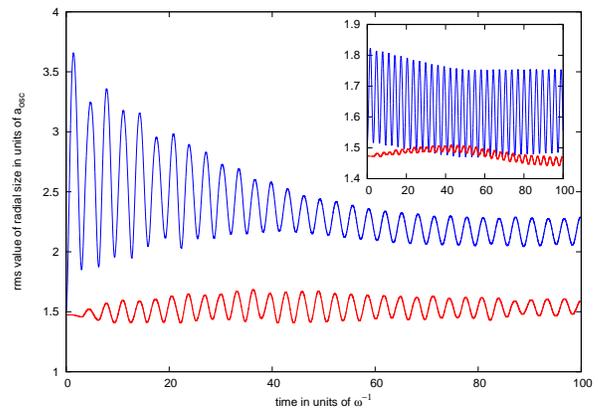}
   \caption{The variation in $r_{\rm rms}$ ( in units of $a_{\rm osc}$ ) for 
            $^{85}$Rb and $^{87}$Rb with time ( in units of $\omega^{-1}$)
            when $a_1$ is changed from $51a_0$ to $408a_0$. The blue and 
            red curves correspond to $^{85}$Rb and $^{87}$Rb respectively.}
 \label{fig-408}
\end{figure}

  When $a_{11}$ is increased to $80a_0$, the $^{85}$Rb condensate oscillates 
radially to accommodate excess repulsion energy. This is the only available 
degree of freedom as tight confinement, arising from  $^{87}$Rb at the flanks, 
along $z$-axis restricts axial oscillations. In TF approximation the effective 
potential $V_{\rm eff}=V + (\mu_2 -V)U_{12}/U_{22}$. The angular frequency 
of the oscillation is $\approx 0.32\omega$. This is close to one of the 
eigen modes of the Bogoliubov equations. The temporal variation of
$r_{\rm rms}$ is shown in Fig.\ref{fig-408} (inset plot). The plots 
show that, the oscillation of the $^{87}$Rb is sympathetically initiated. 
This is due to the coupling between the two condensate species. The 
oscillations are more prominent with less number of atoms.

\begin{figure}[h]
   \includegraphics[width=8cm]{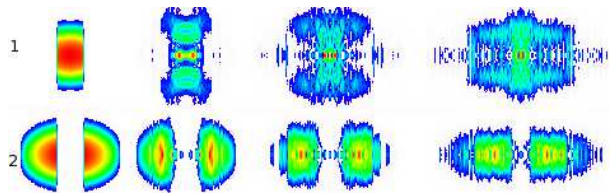}
   \caption{Evolution of the TBEC with RTI. The first and second row are
            density profiles of $^{85}$Rb and $^{87}$Rb BECs respectively
            after increasing $a_{11}$ to $408a_0$. Starting from left, 
            the density profiles are at 0, 24.5, 49.0 and 73.5 msecs 
            after the increase of $a_{11}$. }
   \label{density-408}
\end{figure}

  There is a change in the nature of oscillations when 
$a_{11} > a_{22}(\mu_1 - V)(\mu_2 - V)$. The corresponding stationary state 
has $^{87}$Rb and $^{85}$Rb at the core and flank respectively. The 
$r_{\rm rms}$ oscillation frequency is the same as in $a_{11} < a_{22}$ case. 
But there is a temporal decay of the amplitude till it  equillibrates. The 
decay is due to the expansion of $^{85}$Rb along $z$-axis and is an
unambiguous signature of RTI. The expansion is clearly discernible in the 
density profile as shown in Fig.\ref{density-408} and the rate of decay 
increases with $\Delta a_{11}$. The main plot in Fig.\ref{fig-408} shows 
temporal variation of $r_{\rm rms}$ for $a_{11}=408a_0$, close to the miscible 
domain. There is a strong correlation between the decay rate and nature of 
oscillation. For $a_{11}$ marginally larger than $a_{22}$, the $^{85}$Rb 
condensate tunnels through the $^{87}$Rb condensate. Where as at larger values
the $^{85}$Rb  expands and spreads into the $^{87}$Rb. 

\begin{figure}[h]
   \includegraphics[height=8cm,angle=-90]{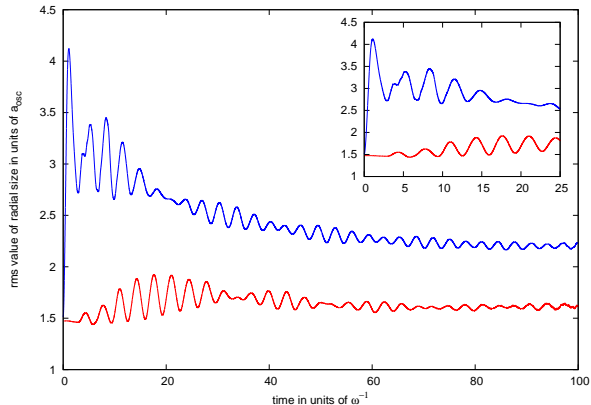}
   \caption{The variation in $r_{\rm rms}$ ( in units of $a_{\rm osc}$ ) for 
            $^{85}$Rb and $^{87}$Rb with time ( in units of $\omega^{-1}$)
            when $a_1$ is changed from $51a_0$ to $780a_0$. The blue and 
            red curves correspond to $^{85}$Rb and $^{87}$Rb respectively.}
   \label{fig-780}
\end{figure}

 A dramatic change of the coupled oscillations occurs when 
$ U_{12}<\sqrt{U_{11}U_{22}} $, the TBEC is then miscible. The $^{85}$Rb 
expands through the $^{87}$Rb cloud and the two species undergo radial 
oscillations which has a beat pattern. The Fig.\ref{fig-780} shows the 
$ r_{\rm rms}$ when $a_{11}=780a_0$. Besides the radial oscillations, as to 
be expected when $a_{11}>a_{22}(\mu_1 - V)/(\mu_2 - V)$, $z_{\rm rms}$ 
increases steadily. This accommodates the excess repulsion energy along the 
axial direction. Along with the oscillations there are higher frequency 
density fluctuations reminiscent of modulational instability. It is to be 
mentioned that, in earlier works \cite{Kasamatsu1,Raju} modulational 
instability in the miscibility domain was analysed in depth. For the present 
case the detailed analysis of modulational instability shall be the subject 
of a future publication.


{\em Summary and outlook}.---We have examined the onset of Rayleigh-Taylor 
instability in TBEC and identified the observable signature in the dynamics. 
We have specifically chosen the experimentally well studied 
$^{85}$Rb-$^{87}$Rb mixture as case study and propose observing RTI with 
the $^{85}$Rb-$^{85}$Rb Feshbach resonance. Starting from $a_{11}< a_{22}$,
RTI sets in when the TBEC is tuned to $a_{11}> a_{22}(\mu_1 - V)/(\mu_2 - V)$
in the TF approximation. Then damping of $r_{\rm rms}$ of $^{85}$Rb, species 
at the core, oscillations marks the onset of RTI. To analyse the stationary 
states we have proposed a semi analytic scheme, applicable when 
$\lambda \ll 1$, to minimize the energy functional with TF approximation. The 
results of which are in excellent agreement with the numerical results.  
The $\lambda \ll 1$ is also the case when the interface is planar and RTI is 
more prominent.


{\em Acknowledgements}.---We thank S. A. Silotri, B. K. Mani and S. 
Chattopadhyay for very useful discussions. We acknowledge the help of 
P. Muruganandam while doing the numerical calculations.


\begin{thebibliography}{99}
\bibitem{asymm} This is a symmetry preserving configuration. The other 
     configuration, the symmetry breaking solution, is energetically not 
     favourable.
\end{thebibliography}
\end{document}